\documentstyle[12pt]{article}
\date{}

\begin{document}
\title{The EPR Argument \\
in a Relational Interpretation \\
of Quantum Mechanics}
\author{Federico Laudisa\\
Department of Philosophy, University of Florence, \\
Via Bolognese 52, 50139 Florence, Italy}
\maketitle
\date{}
\begin{abstract}
\noindent
It is shown that in the Rovelli {\it relational} interpretation of
quantum mechanics, in which the notion of absolute
or observer independent state is rejected, the conclusion of the ordinary
EPR argument turns out to be frame-dependent, provided the conditions
of the original argument are suitably adapted to the new interpretation.
The consequences of this result for the `peaceful coexistence' of quantum mechanics and
special relativity are briefly discussed.
\end{abstract}

\newpage

\section{Introduction}

The controversial nature of observation in quantum mechanics has been at the
heart of the debates on the foundations of the theory since its early days.
Unlike the situation in classical theories, the problem of how and where
we should localize the boundary between systems that observe and systems that
are observed is not merely a practical one, nor has there been a widespread
consensus on whether there is some fundamental difference between the
two classes of systems. The emphasis on the reference to an appropriate
observational context, in order for most properties of quantum mechanical
systems to be meaningful, has been for example the focus of Bohr's reflections
on the foundations of quantum mechanics, and the attention to this
level of description has been inherited to a certain extent even by those
interpretations of the theory that urged to go well beyond the Copenhagen
standpoint.

Although these controversies deal primarily with
the long-studied measurement problem, we are naturally led to ask ourselves
whether a deeper emphasis on the role of the observer might suggest new
directions also about nonlocality.
The celebrated argument of Einstein, Podolsky and Rosen (EPR) for the incompleteness
of quantum mechanics turns essentially on the possibility for an observer of predicting
the result of a measurement performed in a spacetime region that is supposed
to be isolated from the
region where the observer is localized. The quantum mechanical description of a typical
EPR state, however, prevents from conceiving that result as simply revealing a
{\it preexisting} property, so that the upshot of the
argument is the alternative between completeness and locality: by assuming the
former we must give up the latter, and viceversa.
\footnote{Clearly both options are available here since we restrict
our attention to the EPR argument. But after the Bell theorem, it is a
widespread opinion that the only viable option for ordinary quantum mechanics
is in fact the first.}
It then turns out rather natural to ask whether, and to what extent,
taking into due account the fact that quantum predictions are the
predictions {\it of a given observer} affects remarkably the structure and
the significance of the argument.

The ordinary EPR argument is formulated in nonrelativistic quantum mechanics,
whose symmetries constitute a Galilei semigroup.
Therefore, an obvious form of observer dependence that must be taken into account
is the frame dependence that must enter into the description when the whole
framework of the EPR argument is embedded into the spacetime of special
relativity. It can be shown that a relativistic EPR argument still works and,
as a consequence, events pertaining to a quantum system may `depend' on events pertaining
a different quantum system localized in a space-like separated region.
\footnote{Clearly there is no consensus on what a reasonable interpretation of this
`dependence' might be, but a thorough discussion of this point is beyond the scope
of the present paper.}
However, in the relativistic EPR argument, a special attention
must be paid to the limitations placed by this generalization to the attribution of
properties to subsystems of a composite system:
the lesson of a relativistic treatment of the EPR argument lies
in the caution one must take when making assumptions on what an observer
knows about the class of quantum mechanical events taking place in
the absolute elsewhere of
the observer himself. Such observer dependence is essentially linked to
the space-like separation between the regions of the measurements and
prevents from using a property-attributing language without qualifications;
an analysis of these limitations shows that the relativistic EPR argument does not in fact support
the widespread claim that nonlocality involves an {\it instantaneous} creation of properties
at a distance.
\footnote{See (\cite{GG}, \cite{Ghirardi}); for an analysis of the
relevance of this fact to the status of superluminal causation
in quantum mechanics, see \cite{Laudisa}.}

It must be stressed that this argument implies by itself no definite
position about the existence of such influences {\it in the physical
world}: it involves only the {\it logical} compatibility
between the idea of action-at-a-distance and the special relativistic
account of the spacetime structure. More generally, what the argument
is meant to point out is the necessity of a
shift to the relativistic regime, in order
to rigorously assess whether some frequently stated claims about
the metaphysical consequences of
the EPR argument are really consistent with spacetime physics.

But according to a recent {\it relational} interpretation
of quantum mechanics, advanced by Carlo Rovelli (\cite{Rov}), we need not shift to
a fully relativistic quantum theory to find a fundamental form of observer dependence.
In this interpretation, the very notion of state of a physical system
should be considered meaningless unless it is not understood as relative to another
physical system, that plays temporarily the role of observer: when dealing
with concrete physical systems, it is the specification of such observer
that allows the ascription of a state to a system to make sense, so that,
to a certain extent, by a relational point of view the selection of such observers features in
quantum theory as the specification of a frame of reference does in relativity theory.
In the following section the relational interpretation will be
sketched, whereas in section 3 a relational analysis of the
(relativistic) EPR argument will show that whether the ordinary conclusion of the
argument - quantum mechanics is either incomplete or nonlocal - holds or not with respect
to a given observer depends on the frame of reference of the latter.
In the final section we will briefly
discuss the consequences of this analysis for the so called `peaceful
coexistence thesis' concerning the relation between quantum theory and
relativity theory.

\section{Relational quantum mechanics and the observer dependence of states}

In the relational interpretation of quantum mechanics, the relativistic frame dependence
of an observer's predictions is not the only source of observer dependence in quantum mechanics:
a fundamental form of observer dependence is detected already
in {\it non-relativistic} quantum mechanics and that concerns the very
definability of physical state. In the relational
interpretation, the notion of absolute or
observer independent state of a system is rejected: it would make no sense
to talk about a state of a physical system
without referring to an observer, with respect to which that state is defined
(\cite{Rov}). Although the analysis of its assumptions and
its consequences is developed within non-relativistic quantum mechanics,
this claim is somehow reminiscent of the Einsteinian operational critique
of the absolute notion of simultaneity for distant observers, and the main
idea underlying the interpretation is put forward by analyzing the different
accounts that two observers give of the same sequence of events in a typical
quantum measurement process.

Let us consider a system $S$ and a physical quantity
$Q$ that can be measured on $S$. We assume that the possible measurement results are two,
$q_1$ and $q_2$ (for simplicity the spectrum of $Q$ is assumed to be simple
and non-degenerate). The premeasurement state of $S$ at a time $t_1$
can then be written as $\alpha_1\phi_{q_1}+\alpha_2\phi_{q_2},$
with $\alpha ,\beta$ complex numbers
such that $\vert\alpha_1\vert^2+\vert\alpha_2\vert^2=1.$ If we suppose
that a measurement of $Q$ is performed and the measurement result is $q_2,$ according to ordinary
quantum mechanics, at a postmeasurement time $t_2$ the state of $S$ is given
by $\phi_{q_2}.$ Let $O$ be the observer performing
the measurement; the sequence that $O$ observes is then
\begin{equation}
\underbrace{\alpha_1\phi_{q_1}+\alpha_2\phi_{q_2}}_{t_1}\,\,
\Longrightarrow\,\,\,\underbrace{\phi_{q_2}}_{t_2}
\label{eq:sequence-O}
\end{equation}
\noindent
Let us now consider how a second observer $O'$ might describe this same
measurement process, concerning in this case the {\it composite} system
$S+O.$ We denote by $\psi_{init}$ the premeasurement state of $O$ and
by $\psi_{q_1}$ and $\psi_{q_2}$ respectively the
eigenstates of the pointer observable (namely the
states that correspond to recording the $Q$-measurement results
$q_1,q_2$). The premeasurement
state of $S+O$ at $t_1,$ belonging to the tensor product Hilbert space
$\cal{H}_S\otimes\cal{H}_O$
of the Hilbert spaces $\cal{H}_S$ and $\cal{H}_O$ associated to
$S$ and $O$ respectively, is then expressed as
\begin{equation}
\psi_{init}\otimes(\alpha_1\phi_{q_1}+\alpha_2\phi_{q_2});
\label{eq:S+O-t1}
\end{equation}
if $O'$ performs no measurement, by linearity (\ref{eq:S+O-t1}) at $t_2$ becomes
\begin{equation}
\alpha_1\psi_{q_1}\otimes\phi_{q_1}+\alpha_2\psi_{q_2}\otimes\phi_{q_2}.
\label{eq:S+O-t2}
\end{equation}
The measurement process, as described by $O',$ is then given by the sequence
\begin{equation}
\underbrace{\psi_{init}\otimes(\alpha_1\phi_{q_1}+\alpha_2\phi_{q_2})}_{t_1}
\,\,
\Longrightarrow
\,\,
\underbrace{\alpha_1\psi_{q_1}\otimes\phi_{q_1}+\alpha_2\psi_{q_2}\otimes\phi_{q_2}}_{t_2}.
\label{eq:sequence-O'}
\end{equation}
\noindent
So far $O'$ may only claim that the states of $S$ and the states of $O$ are suitably correlated.
But let us suppose now that $O'$, at time $t_3>t_2$ performs on $S$ a measurement of $Q$. Since
we are dealing with the two different observers $O$ and $O'$, it is natural
to ask what general consistency condition we should assume to hold for the different
descriptions that $O$ and $O'$ might give of the measurement process.
In the present situation, such a condition can be the following:
if $S$ at a time $t$ is in an eigenstate $\psi_q$ of an observable $Q$ relative
to $O$, the observer $O'$ that measures $Q$ on $S$ at $t'>t$ (with no
intermediate measurements, between $t$ and $t'$, of observables that are not compatible with
$Q$) will find the eigenvalue belonging to $\psi_q$: therefore
$S$ will be in the state $\psi_q$ also relative to $O'$ (we will return later to the general
form that this consistency condition assumes).
Therefore, the state of $S+O$ at $t_3$ relative to $O'$ will be $\psi_{q_2}\otimes\phi_{q_2}\,,$ since
at $t_2$ the state of $S$ relative to $O$ had been reduced to $\phi_{q_2}.$
If we now compare (\ref{eq:sequence-O}) and (\ref{eq:sequence-O'}), we see that
$O$ and $O'$ give a different account of the same sequence of events
in the above described measurement: at $t_2$ $O$ attributes to $S$
the state $\phi_{q_2},$ whereas $O',$ that views $S$ as a subsystem of
$S+O,$ attributes to $S$ the state $\alpha_1\phi_{q_1}+\alpha_2\phi_{q_2}.$

The most general assumption of the relational interpretation of
quantum mechanics can be then summarized as follows: the
circumstance that there are different ways in which even a simple
quantum mechanical process like the measurement above can be
described by different observers suggests then that this
relational aspect might be not an accident, but a fundamental
property of quantum mechanics. In addition to this, the relational interpretation makes
a pair of further assumptions,
concerning respectively the universality and the completeness of quantum mechanics.

\medskip

\noindent
1. {\it All physical systems are equivalent.}

\noindent
No specific assumption is made concerning the systems that are supposed to act as observers,
except that they must satisfy the laws of quantum mechanics: being observer is not a
property fixed once and for all for a privileged class of physical systems, permanently
identifiable as `observation systems' and clearly separable from the rest of physical systems
(\cite{Rov}, p.1644), nor is it implicitly assumed that such observation systems are
conscious entities.

\smallskip

\noindent
2. {\it Relational quantum mechanics is a complete physical theory.}

\noindent
The general circumstance that different observers may give different
accounts of the same processes is no sign of any fundamental incompleteness
of quantum mechanics, but is simply the consequence of a relational
meta-assumption, according to which there is no `absolute' or `from-outside'
point of view from which we might evaluate the states of a physical system or
the value of quantities measurable on that system in those states.
``Quantum mechanics can therefore be viewed as a theory
about the states of systems and values of physical quantities relative to
other systems. [...] If the notion of observer-independent description
of the world is unphysical, a complete description of the world is exhausted
by the relevant information that systems have about each other.'' (\cite{Rov}, p. 1650).

\medskip

One might be tempted to describe the situation above simply by
saying that the difference between $O$ and $O'$ is that $O$ knows at $t_2$ what the state of $S$
is whereas $O'$ does not, and it is for this reason that $O'$ attributes to $S$ a superposition
state, namely an  informationally `incomplete' state. The problem with this position, however,
is that it implicitly assumes the `absolute' viewpoint on states of physical systems, namely
just the viewpoint that the relational interpretation urges to reject as implausible.
Moreover, on the basis of the above account, one might draw a general distinction
between
a {\it description} and an {\it observation} of a system $S$ by an observer $O$:
in the former case, a `description' of $S$ involves no interaction between
$O$ and $S$ {\it at the time in which $O$ describes $S,$} although it is
still necessarily based on some {\it prior} interaction between $S$ and
other systems.
On the other hand, we may say that $O$ `observes' $S$ when $O$ actually
measures some relevant physical quantity on $S$: it is clear that,
in this case, there is an interaction between $S$ and $O$ that occurs
exactly when $O$ is said to `observe' $S.$

If we return to our specific example, we might consider the
`description' that $O'$ gives of $S$ {\it at a given $t$} in terms of
correlation properties of the system $O+S$ as the maximal amount of information on
the measurement process involving $S$ and $O$ that is available to $O'$
in absence of interaction {\it at $t$} between $O'$ and the composite system
$O+S.$ So let us recall the sequence (\ref{eq:sequence-O'}) of events.
$O'$ is supposed to perform no measurement in the time interval $[t_1,t_2]$
and thus can `describe' the state of $S$ at $t_2$ only through some observable
$C_{(O,S)}$ (defined on $\cal{H}_S\otimes\cal{H}_O$), that tests whether $O$ has
correctly recorded the result of
the measurement on $S.$ The eigenvalues of $C_{(O,S)}$ are simply $1$ and $0$.
The states $\psi_{q_1}\otimes\phi_{q_1},$ $\psi_{q_2}\otimes\phi_{q_2}$
turn out to be eigenstates belonging to the eigenvalue $1$ - the
record was correct - whereas the states
$\psi_{q_1}\otimes\phi_{q_2},$ $\psi_{q_2}\otimes\phi_{q_1}$
turn out to be eigenstates belonging to the eigenvalue $0$ - the
record was incorrect, namely
\begin{eqnarray*}
C_{(O,S)}\,\psi_{q_1}\otimes\phi_{q_1}=\psi_{q_1}\otimes\phi_{q_1}, & & C_{(O,S)}\psi_{q_1}\otimes\phi_{q_2} = 0\\
C_{(O,S)}\,\psi_{q_2}\otimes\phi_{q_2}=\psi_{q_2}\otimes\phi_{q_2}, & & C_{(O,S)}\psi_{q_2}\otimes\phi_{q_1}=0.\\
\end{eqnarray*}
On the basis of the above account, the consistency
requirement that we mentioned earlier, and that concerns the relation between
different descriptions of the same event given by different
observers, appears rather natural: it can be expressed as the requirement that if
the only information available to $O'$ is that $O$ has measured $Q$ on $S$
but the result is unknown, the results that $O'$ obtains by performing
a $C_{(O,S)}$-measurement and a $Q$-measurement must be correlated  (\cite{Rov}, pp. 1650-2).

\section{A relational analysis of the EPR argument}

An ideal place to look at to see how a relational approach might change the
`absolute' view of quantum mechanical states is just
the framework in which the EPR argument is usually
developed (\cite{EPR}, \cite{Bohm}).

The physical framework common to all variants of this argument involves
a two-particles' system, whose subsystems interact for
a short time and then separate.
The original formulation of the EPR argument takes into account
a pair of quantities for each
particle, such in a way that the members of each pair are
mutually incompatible.
We will take into account the usual non-relativistic Bohm version of
the original argument, formulated as a spin correlation experiment,
and in a simplified form that deals with just one quantity for each particle (\cite{Redhead}).
The assumptions of the argument will be slightly rephrased as compared to the widespread
formulation, but this rephrasing does not substantially affect the argument.

\begin{enumerate}

\item {\sc Reality}

\noindent
If, without interacting with a physical system $S,$ we can predict
with certainty - or with probability one - the result $q$ of
the measurement of a quantity $Q$ performed at time $t$ on
$S,$ then at a time $t'$ immediately after $t,$ there exists a
property - associated with $q$ and denoted by $[q]$ - that is actually
satisfied by $S$: any such property is said to be an {\it objective} property
of $S.$

\medskip

\item {\sc Completeness}

\noindent
Any physical theory $T$ describing a physical system $S$ accounts for
every objective property of $S.$

\medskip

\item {\sc Locality}

\noindent
No objective property of a physical system $S$ in a state $s$
can be influenced by measurements performed at a distance on a different
physical system.

\medskip

\item {\sc Adequacy}

\noindent
The statistical predictions of quantum mechanics are correct.
\end{enumerate}

\medskip

\noindent
A word of comment on the formulation of the Reality condition is in order.
As is expected, the notion of an objective property $[q]$ of
a physical system $S$ is equivalent to the notion of $S$ having $[q]$
no matter whether $Q$ is measured on $S$ or not. Clearly $S$ may
have non-objective properties such as, for instance, `correlation'
properties: certain possible values of a quantity $R$ measurable on $S$ are
correlated to possible values of a quantity pertaining the pointer
of an apparatus that is supposed to measure $R$ on $S.$ Obviously $S$
cannot be said to have such properties independently from any
measurement (actual or not) of $Q.$

The experimental situation considered (often called
an EPR-Bohm correlation experiment) involves a two spin-1/2 particles'
system $S_1+S_2$ prepared at the source in the singlet state. If we focus only on the spin part,
such state of $S_1+S_2$ can be written for any spatial direction $x$ as
\begin{equation}
{1\over\sqrt 2}[\phi_{1,x}(+)\otimes\phi_{2,x}(-)\,-\,
\phi_{1,x}(-)\otimes\phi_{2,x}(+)]\,,
\label{eq:singlet}
\end{equation}
where:
\begin{itemize}
\item
$\phi_{i,x}(\pm)$ is the eigenvector of the
operator $\sigma_{i,x},$ representing spin up or down along the direction
$x$ for the particle $i=1,2;$
\item
$\phi_{1,x}(+)\otimes\phi_{2,x}(-)$ and
$\phi_{1,x}(-)\otimes\phi_{2,x}(+)$ belong to the tensor product
${\cal H}_1\otimes{\cal H}_2$ of the Hilbert spaces ${\cal H}_1$ and
${\cal H}_2$ associated respectively to subsystems $S_1$ and $S_2$.
\end{itemize}
$S_1$ and $S_2$ are
supposed to fly off in opposite directions; spin measurements are
supposed to be performed when $S_1$ and $S_2$ occupy two widely separated
spacetime regions $R_1$ and $R_2,$ respectively.
It follows from (\ref{eq:singlet}) and Adequacy that if on measurement
we find spin up along the direction $x$ for the particle $S_1,$
the probability of finding spin down along
the same direction $x$ for the particle $S_2$
equals $1$: it is usual to say that $S_1$ and $S_2$ are strictly
anticorrelated.

Let us suppose now that, for a given direction $z,$
we measure $\sigma_{1,z}$ at time $t_1$ and we find $-1.$ Adequacy then allows
us - via anticorrelation - to predict with probability one the result of
the measurement of $\sigma_{2,z}$ for any time $t_2$ immediately after
$t_1,$ namely $+1.$ Then, according to Reality,
there exists an objective property $[+1]$ of $S_2$ at
$t_2.$ By Locality, $[+1]$ was an objective property of $S_2$ also
at a time $t_0 < t_1,$ since otherwise it would have been
`created' instantaneously by the act of performing
a spin measurement on $S_1.$ However, at time $t_0$
the state of $S_2$ is  a mixture, namely ${1\over 2}(P_{\phi_{2,x}(+)}+ P_{\phi_{2,x}(-)})$,
since the entangled state
(\ref{eq:singlet}), although it is a pure state of $S_1+S_2,$ uniquely
determines the states of the subsystems as mixed states.
Thus $S_2$ is shown
to satisfy an objective property in a state which is not an eigenstate of
$\sigma_{2,z}.$ However all quantum mechanics is able to predict is the
satisfaction of properties such as $[+1]$
only in eigenstates of $\sigma_{2,z}:$ quantum mechanics then
turns out to be incomplete, since there exist objective properties that are
provably satisfied by a system described by quantum mechanics but that cannot
be described in quantum mechanical terms.
By a strictly logical point of view, the conclusion of the argument can
be rephrased
as the statement that the conjunction of Reality, Completeness, Locality
and Adequacy leads to a contradiction.

In the framework of ordinary
quantum mechanics, Reality and Adequacy cannot be called into question:
whereas the latter simply assumes that the probabilistic statements of
quantum mechanics are reliable,
without the former no quantum system could ever satisfy objective
properties, not even such
a property as having a certain value for a given quantity
one is going to measure, when
the system is prepared in an eigenstate belonging to that (eigen)value
of that quantity.
\footnote{See for instance the clear discussion in \cite{BLM}
on objectivity and non-objectivity of properties in quantum mechanics.}
The alternative then reduces to the choice between Completeness and
Locality: by assuming Completeness, we then turn the above EPR argument into
a nonlocality argument. A relativistic formulation of this argument can also be given:
although the different geometry of spacetime must be taken into account, the only generalization
lies in adapting the Locality condition, to the effect that
objective properties of a physical system cannot be influenced by measurements performed
in space-like separated regions on a different physical system (\cite{GG}).

Let us turn now to a relational analysis of the argument.
In a relational approach to the EPR argument, we have to modify accordingly
its main conditions (Adequacy is obvious), basically by relativizing the objectivity of
properties to given observers. The new versions might read as follows:

\begin{enumerate}

\item[$1'.$] {\sc Reality$^*$}

\noindent
If an observer $O$, without interacting with a physical system $S,$
can predict with certainty (or at any rate with probability one)
at time $t$ the value $q$ of a physical quantity $Q$
measurable on $S$ in a state $s,$ then, at a time $t'$ immediately after $t,$
$q$ corresponds to a property of $S$ that is objective relative to $O.$

\item[$2'.$] {\sc Completeness$^*$}

\noindent
Any physical theory $T$ describing a physical system $S$ accounts for
every property of $S$ that is objective relative to some observer.

\item[$3'.$] {\sc R-Locality$^*$}

\noindent
No property of a physical system $S$ that is objective relative to some
observer can be influenced by measurements performed in space-like separated regions
on a different physical system.
\end{enumerate}

\noindent
Once the above weaker sense of objectivity is defined, Completeness$^*$ is little more
than rephrased, whereas R-Locality$^*$ guarantees that no
property that is non-objective (in the weaker sense of objective as relative to a given observer) can be
turned into an objective one (still in the weaker sense) simply by means of operations
performed in space-like separated regions (at this stage, already the relativistic version
of the Locality condition in the ordinary EPR argument is assumed).

So far the relational versions of the original EPR conditions have
been introduced: but are they sufficient in order to derive the
same conclusion drawn by the original argument? We are interested
into the states of the subsystems $S_1$ and $S_2$ and the values
of spin in those states when $S_1$ and $S_2,$ that initially
interact for a short time and then fly off in opposite directions,
are localized in space-like separated regions $R_1$ and $R_2$
respectively. According to the relational interpretation the
reference to the state of a physical system is meaningful only
relative to some observer. So let us suppose to introduce two
observers, $O_1$ for $S_2$ and $O_2$ for $S_2,$ that, after
coexisting in the spacetime region of the source for the short
time of the interaction, follow each its respective subsystem.
Initially, at time $t_0$, $O_1$ and $O_2$ agree on the state
(\ref{eq:singlet}). But when, after leaving the source, $S_1$ and
$S_2$ are subject to measurement, the spacetime regions in which
the measurements are supposed to take place are mutually isolated.
Then we suppose that, for a given direction $z,$ an observer $O_1$
measures $\sigma_{1,z}$ on $S_1$ at time $t_1 > t_0$ and finds
$-1.$ Now the strict spin value anticorrelation built into the
state (\ref{eq:singlet}) allows $O_1$ to predict with certainty
the spin value for $O_2$ without interacting with it. Namely
Adequacy allows $O_1$ to predict with probability one the value of
$\sigma_{2,z}$ on $S_2$ for any time $t_2$ immediately after
$t_1,$ namely $+1.$ Then, according to Reality$^*$, there exists a
property $[+1]$ of $S_2$ that is objective {\it relative to} $O_1$
at $t_2.$ By R-Locality$^*$, $[+1]$ was an objective property of
$S_2$ relative to $O_1$ also at a time $t_0 < t_1,$ since
otherwise it would have been created by the act of
performing a spin measurement on the space-like separated system
$S_1.$ Then $O_1$ may backtrack at $t_0$ the property $[+1]$ or,
equivalently, the {\it pure} state $\phi_{2,z}(+).$ At this point
the non-relational argument proceeds by pointing out that at $t_0$
the state of $S_2$ as determined by (\ref{eq:singlet}) was the
mixture ${1\over 2}(P_{\phi_{2,x}(+)}+ P_{\phi_{2,x}(-)})$: since
this kind of state cannot possibly display a property like $[+1],$
the charge of incompleteness for quantum mechanics follows. By a
relational viewpoint, however, this conclusion does {\it not}
follow: in fact, we are comparing two states relative to {\it
different} observers. Since the regions $R_1$ and $R_2$ of the
measurements are space-like separated, there are frames in which
the measurement performed by $O_1$ precedes the measurement
performed by $O_2$. Let us suppose that in one of these frames
$O_1$ performs a spin measurement on $S_1$ along the direction $z$
at a time $t_1$ and finds $-1$; then, at $t_2$ immediately after
$t_1$, $S_1$ is in the state $\phi_{1,z}(-)$ relative to $O_1$,
and the latter is allowed to attribute to $S_2$ the pure state
$\phi_{2,z}(+)$ at $t_2$, corresponding to a spin value $+1$.
Thus there is a property $[+1]$ of $S_2$ that is objective relative
to $O_1$; hence, according to R-Locality$^*$, $O_1$ should be
allowed to backtrack at $t_0<t_1$ that same property $[+1]$ - satisfied by
$S_2$ in $\phi_{2,z}(+)$ - but in this case he would derive
incompleteness, since at that time the state of $S_2$ is a
mixture. At time $t_0$, however, $O_2$ did not perform yet any
spin measurement: therefore $O_2$ is still allowed to attribute to $S_1$
and $S_2$ only the mixtures ${1\over 2}(P_{\phi_{1,x}(+)}+
P_{\phi_{1,x}(-)})$ and ${1\over 2}(P_{\phi_{2,x}(+)}+
P_{\phi_{2,x}(-)})$ respectively, and there is no matter of fact
as to which observer `is right'. Until $O_2$ does not perform a
measurement, $O_2$ may then describe the measurement by $O_1$
simply as the establishment of a correlation between $O_1$ and
$S_1$.

The conclusion to be drawn is then that the question:
{\it when is an observer allowed to claim, via the EPR argument, that quantum mechanics is either incomplete
or nonlocal?} has a frame-dependent answer.
Let us denote with $M_O^t(\sigma, S,\rho,R)$ the event that an
observer $O$ performs at time $t$ a measurement of a physical quantity $\sigma$ on a system
$S$ in the state $\rho$ in the region $R$.
If we set
\begin{eqnarray*}
M_R & := & M_{O_1}^{t^R}(\sigma_{1,x}, S_1,{1\over 2}(P_{\phi_{1,x}(+)}+
P_{\phi_{1,x}(-)}),R_1)\\
M_L & := & M_{O_2}^{t^L}(\sigma_{2,x}, S_2,{1\over
2}(P_{\phi_{2,x}(+)}+ P_{\phi_{2,x}(-)}),R_2),
\end{eqnarray*}
the three possible cases are:

\noindent
(a) $M_R$ precedes $M_L$;

\noindent
(b) $M_R$ follows $M_L$.

\noindent
(c) $M_R$ and $M_L$ are simultaneous.

In case (a), the EPR argument works for $O_1$ but not for $O_2$, namely
quantum mechanics is either incomplete or nonlocal relative to
$O_1$ but not relative to $O_2$, whereas the situation is reversed in the case (b).
Finally, in the case (c), what the EPR argument implies is nothing but the outcome-outcome
dependence built into the correlations intrinsic to EPR entangled states (\cite{Shimony1}).
However, until $O_1$ and $O_2$ do not interact, they cannot compare their respective state
attributions so that, up to the interaction time, again each observer can claim either
the incompleteness or the nonlocality of quantum mechanics only relative to himself.

\section{Final discussion}

As we recalled above, the EPR incompleteness argument for ordinary quantum
mechanics can be turned into a nonlocality argument, that appears to
threaten the mutual compatibility at a fundamental level of quantum theory
and relativity theory. The received view has been that, relevant that
nonlocality may be by a foundational viewpoint, the conflict engendered by
it is not as deep as it seems. There would be in fact a `peaceful
coexistence' between the two theories, since locality in quantum mechanics
would be recovered at the statistical level and in any case - it is argued -
nonlocal correlations are uncontrollable so no superluminal transmission of information
is allowed.
\footnote{See for instance \cite{Eberhard}, \cite{GRW} and \cite{Shimony1}.
For dissenting views, one can see \cite{CJ} and \cite{Muller}.}

Such view, however, has several drawbacks, first of all the fact that it is based
on highly controversial notions such as the `controllability' (or
`uncontrollability') of information, or on the vagueness of establishing when
it is exactly that an `influence' becomes a bit of information, or
on whether relativity theory prohibits superluminal exchanges of information but not
superluminal travels of influences, and subtleties
like that. On the other hand, the very inapplicability of
the ordinary EPR argument in relational quantum mechanics prevents this kind of
controversies and provides a new way to interpret the peaceful coexistence
thesis, since in the relational interpretation no compelling alternative between completeness
and locality via the EPR argument can be derived in a frame-independent way.

The observer dependence that affects the EPR argument in a relational approach to
quantum mechanics might also imply a different view of the hidden variables program.
Most hidden variable theories do essentially two things. First,
on the basis of the EPR argument's conclusion they {\it assume}
the incompleteness of quantum mechanics as a starting point; second, they introduce
the hypothesis of a set of `complete' states of a classical sort, the averaging
on which gives predictions that are supposed to agree with the quantum mechanical ones
(see for instance \cite{CS}, \cite{Redhead}).
If however, in a relational approach to quantum mechanics, the conditions of Reality$^*$,
Completeness$^*$, R-Locality$^*$ and Adequacy need not clash from
one frame of reference to the other, the attempt of `completing' quantum mechanics by
introducing hidden variables turns out to be unmotivated by a relational point of view,
and so does any nonlocality argument that has this attempt as a premise. Moreover, the
hypothetical complete states of hidden variables theories are conceived themselves as
observer-{\it in}dependent, so that an absolute view of the states of physical system
would be reintroduced, albeit at the level of the hidden variables.

The issue of whether statistical correlations across space-like
separated regions are a real threat to a peaceful coexistence between quantum theory
and relativity theory is being more and more investigated also in the framework of algebraic
quantum field theory (AQFT): in fact a suitable form of Bell inequality -
an inequality that in different formulations has been extensively
studied with reference to the nonlocality issue in
non-relativistic quantum mechanics (\cite{Bell}) - has been shown
also to be violated in AQFT (see for instance \cite{SW} and
\cite{S}). The questions of whether a relational interpretation of
AQFT might be developed, and of what such interpretation might
have to say about the violation of the Bell inequaliity in AQFT,
appear worth investigating for several reasons. First, an
absolute view of quantum states seems in principle precluded in
AQFT. In the algebraic framework for quantum field theory,
local algebras are axiomatically associated with specific (open bounded) regions
of the Minkowski spacetime: the elements of the algebras are represented as the observables
that can be measured in the spacetime region to which the algebra is associated. The states,
represented as suitable expectation functionals on the given algebra, encode the statistics
of all possible measurements of the observables in the algebra and thus inherit from
the latter the feature of being defined not globally but with respect to a particular
spacetime region. Second, the derivation of a suitable
Bell inequality and the analysis of its violation refer neither to
unspecified hidden variables nor to the need of introducing them.
Finally, the locality that the violation of the suitable Bell
inequality might call into question is
directly motivated by relativistic constraints and is
not a hypothetical condition satisfied only by hidden variables,
and imposed over and above a theory that is intrinsically nonlocal
such as ordinary quantum mechanics.

\bigskip

\noindent
{\bf Acknowledgements}

\medskip

\noindent
I am deeply grateful to Carlo Rovelli for an interesting
correspondence on an earlier version of this paper:
his insightful comments greatly helped me to clarify a number of relevant
points, although I am of course the only responsible of the arguments defended here.
This work was completed during a visit to the Department of History and Philosophy
of Science of the E\"otv\"os University in Budapest. I wish to thank in particular
Mikl\'os R\'edei and L\'aszl\'o Szab\'o for their support and their hospitality.
This work is supported by a NATO CNR Outreach Fellowship.

\end{document}